# A Numerical Method to Compute Brain Injury Associated with Concussion

C. Bastien[1], A. Scattina[2], C. Neal-Sturgess[3], R. Panno[2], V. Shrinivas[1].

[1]*Coventry University, Institute for Future Transports and Cities, Coventry, UK*
[2] *Politecnico di Torino, Department of Mechanical and Aerospace Engineering, IT*
[3]*University of Birmingham, Department of Mechanical Engineering, Birmingham, UK*

**Abstract:** Concussion can result from various events in everyday life, including falls, sports collisions, and motor vehicle accidents, which could lead to the disruption of neuronal cell membranes and axonal stretching, leading to a neuro-metabolic cascade of molecular changes in the brain. There is currently no agreement on which computational method can assess such low-level injuries. This paper demonstrates for the first time that the Peak Virtual Power (PVP), based on the Clausius-Duhem inequality, assuming that the injury is represented by the irreversible work in a human body, could be a candidate to capture brain distortion related to concussion. The work is based on the evaluation of the PVP via reconstruction of three NFL helmet-to-helmet impacts by means of finite element analysis, using validated Biocore helmet models fitted with calibrated Hybrid III headforms against linear and angular acceleration impact corridors, which were defined as realistic impact conditions for each collision scenario. Once the exact impact parameters were defined, the Hybrid III headform was replaced with a validated THUMS 4.02 human head model in which the PVP was computed for each head at the corpus callosum and midbrain locations. The results indicate that mild and severe concussions could be prevented for lateral collisions and frontal impacts with PVP values lower than 0.928mW and 9.405mW, respectively, and no concussion would happen in the head vertical impact direction for a PVP value of less than 1.184mW. This innovative method proposes a new paradigm to improve helmet designs, assess sports injuries and improve people's wellbeing.

## Highlights

- Peak Virtual Power method can capture brain distortion related to concussion
- Concussion is extracted from corpus callosum and midbrain locations of THUMS4.02
- Peak power in midbrain less than 1.184mW for a vertical impact leads to no concussion
- Peak power in midbrain more than 0.928mW for a lateral impact leads to concussion
- Peak power in midbrain more than 9.405mW for a front impact leads to concussion

## 1 Introduction

Concussion can result from various events in everyday life including falls, sports collisions and motor vehicle accidents. An impact to the head could disrupt neuronal cell membranes, leading to a neuro-metabolic cascade of molecular changes in the brain, which increases vulnerability to a repeat injury [1]. A high incidence of concussions has been reported in sports: there are approximately 3.8 million sports-related concussions per year in the USA [2] and it has been estimated that about 19% of participants in contact sports, such as American football, suffer at least one concussion during a competitive season [3]. However, its recognition is still difficult and concussive events often go untreated. In recent years, the correlation between repeated concussive events and the development of diseases such as dementia, depression, Alzheimer's and chronic traumatic encephalopathy (CTE) is becoming increasingly evident, after studies conducted on the brains of retired players of the National Football League (NFL) [4][5]. Hence, this underlines the importance of improving concussion recognition and developing effective strategies against repeated injury in sports.

In contact sports, the majority of concussive events occur after player-to-player collisions, in particular head-to-head impacts resulting in the highest rate of concussions. In American football head-to-head contact takes place when the helmets of two players strikes with a high degree of force. Intentionally causing such types of collisions is banned in most leagues, due to the elevated risk of injury it presents.

In the literature two different kinds of parameters are analysed in order to evaluate concussion:

- Global kinematic parameters: maximum linear and rotational acceleration and their duration;

- Intracerebral parameters: maximum axonal strain, maximum strain energy, maximum von Mises stress, maximum von Mises strain, maximum shear stress, maximum shear strain, maximum principal stress, maximum principal strain, minimum and maximum pressure [6][7][8][9].

However, these parameters can only approximate a mechanical threshold for concussion (often not even unambiguous), estimating a tolerance level for a 50% risk of injury [6][7][8][9] when used in finite element human computer models. Furthermore, using strains is not useful to capture the injury severity range, as well as trauma location, which has been proven to be too diffused to be of any relevance [10]. The Peak Virtual Power (PVP) method, on the other hand, can overcome these shortcomings. The PVP is derived from the rate-dependent form of the 2nd law of thermodynamics using the Clausius-Duhem inequality, assuming that the injury is represented by the irreversible work in a human body [10]. This method is scientifically rigorous and based on theoretical physics (Equation 1) and has been proven to compute accurately the severity of brain white and grey matter injury, and more particularly the location of the brain damage [10][11]

$$PVP \propto max(\sigma \cdot \dot{\varepsilon}) \propto AIS$$

Equation 1: Peak Virtual Power. Relationship between PVP and AIS [14]

The exact geometry of the rat brain was recreated [12], and impacts were recorded in 5 segments illustrated in Figure 1, measuring strain and strain rate in each segment. Figure 1 highlights that segments 2 and 3 show the highest brain distortion.

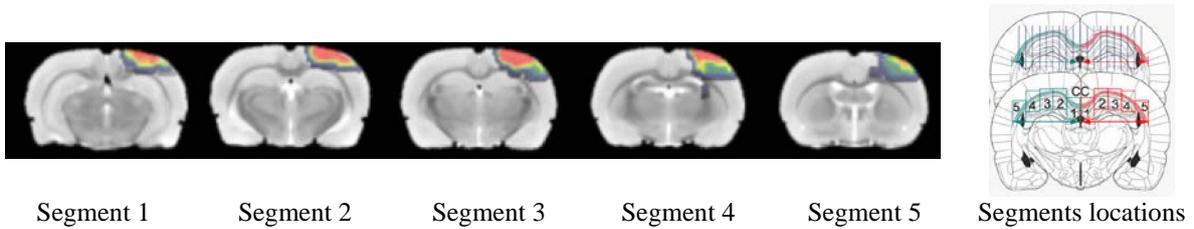

    Segment 1    Segment 2    Segment 3    Segment 4    Segment 5    Segments locations

Figure 1: Rat brain against mild impact, showing contusion/oedema (Imaging)

The maximum strain and strain rates for each segment of that study were recorded in Figure 2.

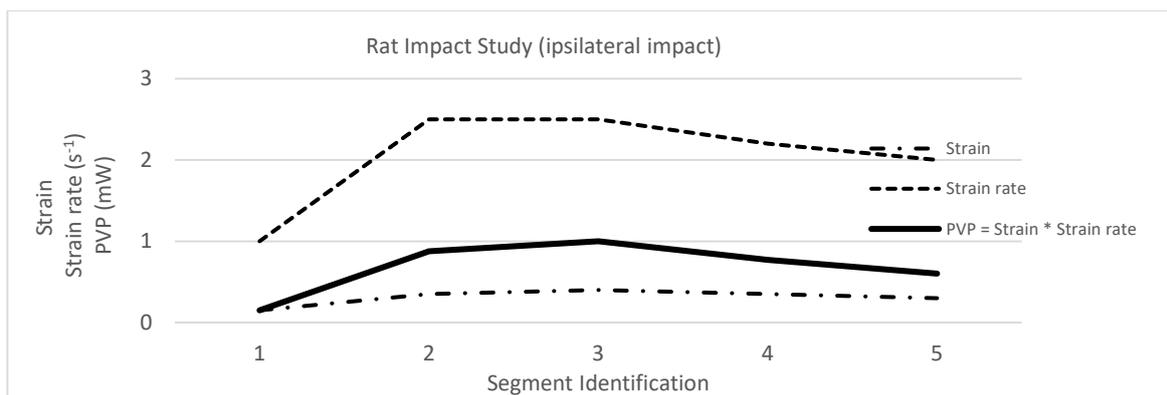

Figure 2: Rat Brain impact study results [12]

It is impossible to determine which part of the brain is more distorted when strain and strain rate are considered independently. However, because stress and strain are proportionally related to Young's modulus, the strain and strain-rate numbers for each section can be compounded to represent the PVP (Equation 1). According to the entity "strain * strain rate" plotted in Figure 2, which is consistent with Figure 1, which demonstrates that PVP could be used to estimate maximum brain deformation, segments 2 and 3 are likely to sustain the most damage. Therefore, PVP will be used in this study to assess brain deformation that results in concussion. In this paper, we will propose a PVP tolerance for each impact direction that causes mild concussion.

## 2 Methodology

This paper will calculate the PVP in the brain centre in real concussive head-to-head impacts in American Football. To measure and analyse the effects of the impact on the brain, the THUMS FE model Version 4.02 AM50 was chosen because of the-detailed modelling of internal organs compared to previous versions [13]. For the purpose of this paper, only the head with its associated organs are extracted from the complete THUMS model.

The methodology will be based on the following steps:

1. Selection of American football collisions concussion cases
2. Calibrate the American football collisions against reconstruction data on the Hybrid III headform
3. Replace the Hybrid III headform for each collision with the THUMS 4.02 human head model
4. Compute the PVP values for each collision
5. Propose a PVP threshold to describe concussion

### 2.1 Selection of American football collisions concussion cases

The data from the laboratory reconstruction of 3 out of 31 incidents of helmet-to-helmet impact resulting in mild concussion during National Football League games was used to replicate real concussive impacts in virtual simulations [14][15][16]. The concussions have been rated as "mild" as the players carried on playing the game. The experimental tests were carried out using helmeted 50th percentile adult male Hybrid III rigid dummies (head-neck complex), simulating the struck and the striking players, with the same impact velocity, direction, and head kinematics as in the game. The kinematic parameters were determined by a video analysis of the impact. The Hybrid III heads were instrumented with nine linear accelerometers to measure both linear and rotational accelerations [14]. The Hybrid III heads were equipped with Riddell VSR-4 helmets. A helmeted head/neck assembly (simulating the struck player) was guided in freefall from a height sufficient to achieve the same impact velocity as that determined from the video analysis of the game impact. Impact was against another helmeted head/neck assembly (simulating the striking player) attached to a freely suspended anthropometric test device torso or against a simulated ground surface [17] (Figure 3).

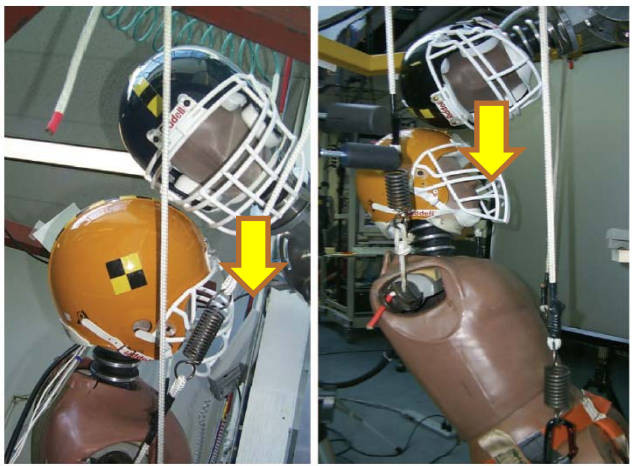

Figure 3: Laboratory set-up. Equipment used during laboratory tests to reconstruct game impacts [17]

In order to compare the effects of each alternative helmet-to-helmet condition on the struck player (Figure 4), the cases numbered 38, 77, and 164 in [15][16] where reconstructed as FE models with parameters from Table 1.

| Case | Initial Velocity (m/s) | Player (Condition) | Linear Acceleration Peak (g) | Rotational Acceleration Peak ($rad/s^2$) | HIC | Mild Concussion |
|---|---|---|---|---|---|---|
| 38 | 9.5 | Struck | 118.5 | 9678 | 554 | Yes |
|  |  | Striking | 59.9 | 5205 | 127 | No |
| 77 | 9.9 | Struck | 80.3 | 5148 | 185 | Yes |
|  |  | Striking | 34.9 | 2714 | 53 | No |
| 164 | 10.8 | Struck | 123.7 | 9590 | 370 | Yes |
|  |  | Striking | 88.7 | 6136 | 202 | No |

Table 1: Impact location and reconstruction results for the struck (concussed) and the striking players

### 2.2   Calibration Process of Concussion Cases

The kinematics of the FE head models were compared and verified with the kinematics corridors derived from the laboratory tests. This process was repeated for all three cases analysed. The 50th percentile adult male Hybrid III head-neck complex model fitted with modified Riddell Revolution Speed Classic Helmet Model v2.0 from Biocore LLC [18] were used in the FE modelling.

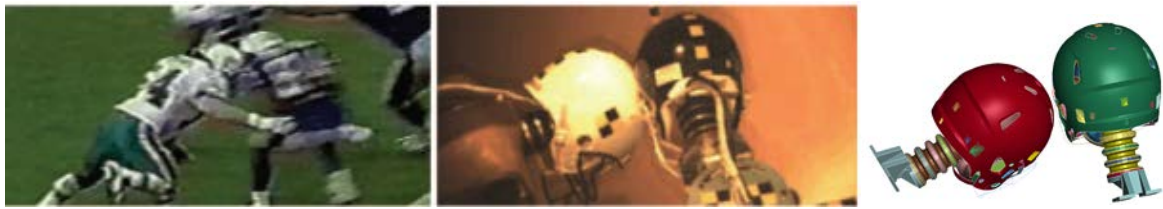

**Figure 4: Impact reconstruction:** From game video analysis (on the left) to laboratory re-enactments (centre) to virtual simulation (right) of a real concussive impact in American Football [14]

### 2.3   Replacement of Hybrid III head model by THUMS 4.02 head for concussion assessment

To assess the effects on the players' brains in the reconstructed impacts, the Hybrid III rigid dummy were replaced with the heads of the THUMS model, as shown in Figure 5. Great care was taken to avoid and remove any initial penetrations and contact crossed edges between the human head and the helmet, due to the geometrical differences between the ATD and the human head.

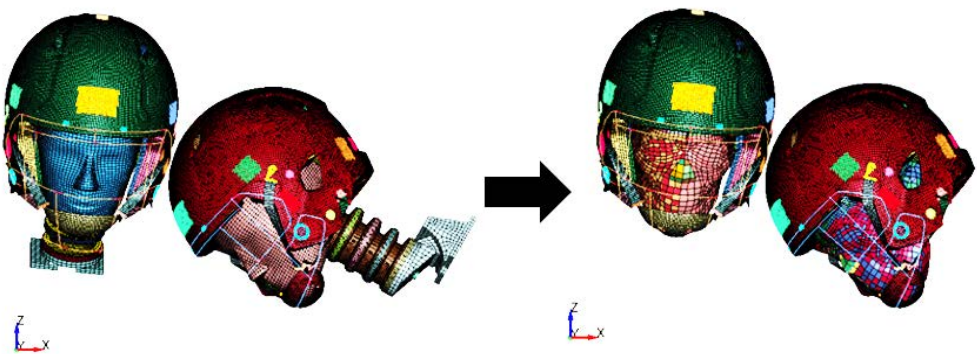

Figure 5: Replacement of the Hybrid III FE model with the THUMS FE model. Case 38: Hybrid III FE model on the left and THUMS FE model on the right.

## 2.4 Computation of Concussion using PVP

A new part division of the brain was defined, starting from the white matter of the THUMS FE model, enclosing the regions of the brain where concussion is initiated, i.e. the corpus callosum and midbrain [19][20][21], as shown in Figure 6. This division was necessary to evaluate concussion for this specific area of the brain.

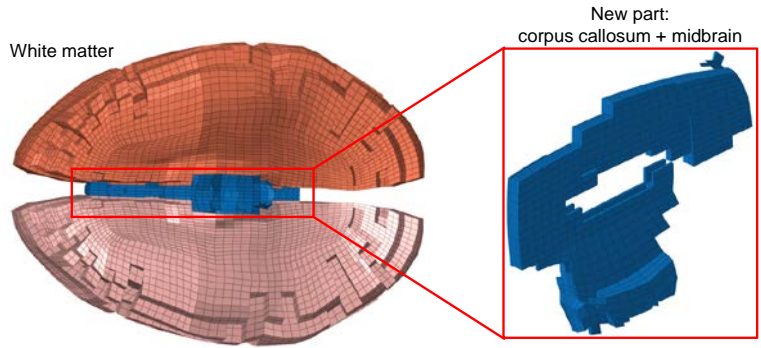

Figure 6: New division extracted from the White Matter of the THUMS FE model.

## 3 Results

### 3.1 Results of Hybrid III correlation

The three collisions have been simulated using LS-Dyna v13.1.0. The relative position of each head has been parametrised into *INCLUDE_TRANSLATE and *INCLUDE_ROTATE commands. The parametric model was fed into the optimiser LS-OPT to define the initial position of the three heads in the three cases that best fit the experimental results in terms of linear and angular acceleration. The initial impact positions of the three cases are illustrated in Figure 7.

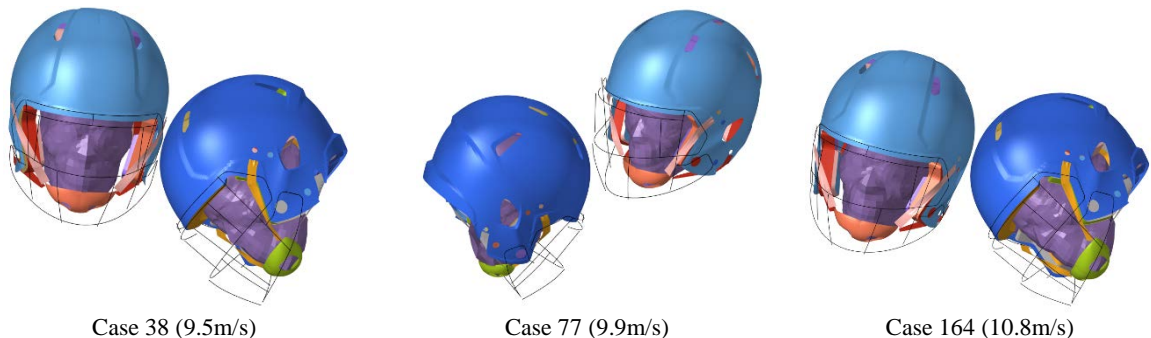

Case 38 (9.5m/s)     Case 77 (9.9m/s)     Case 164 (10.8m/s)

Figure 7: Actual head positions optimised for each collision case

The linear and angular acceleration correlation results (plotted with standard CFC1000 cut-off filters) are provided in Figure 8 and Figure 9 for case 38, Figure 10 and Figure 11 for case 77, Figure 12 and Figure 13 for case 164.

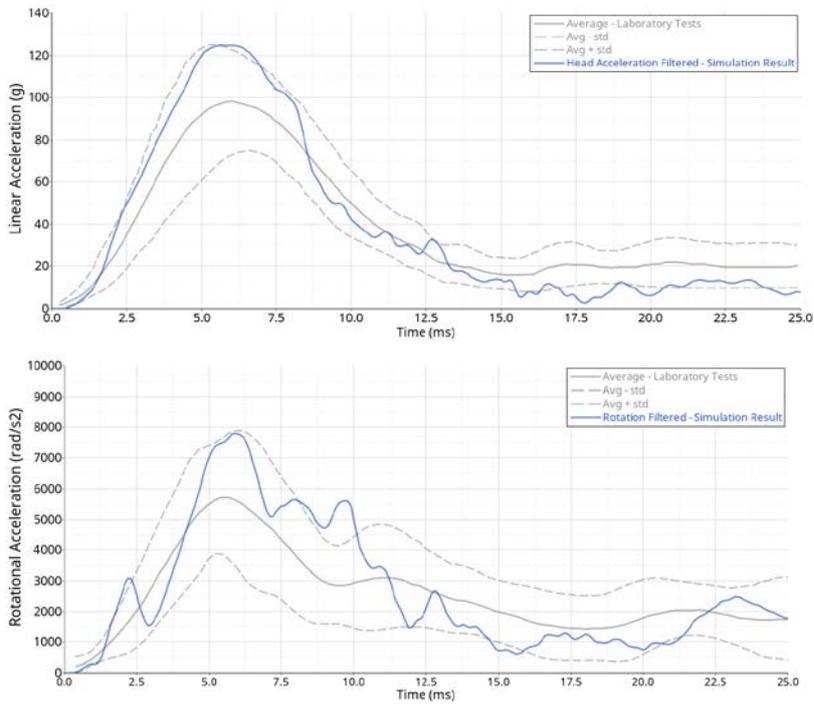

Figure 8: Case 38 - Struck player Hybrid III model response vs tests

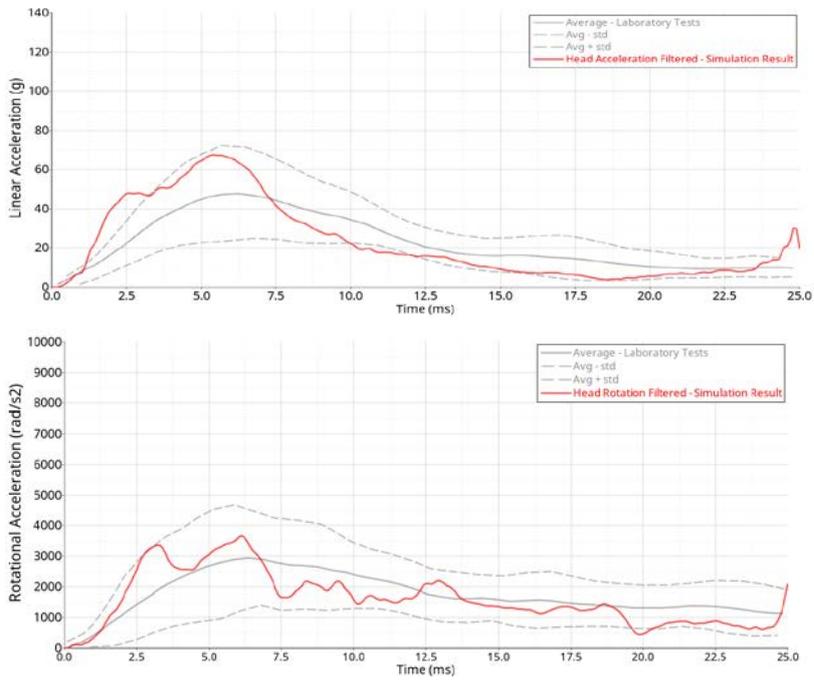

Figure 9: Case 38 - Striking player Hybrid III model response vs tests

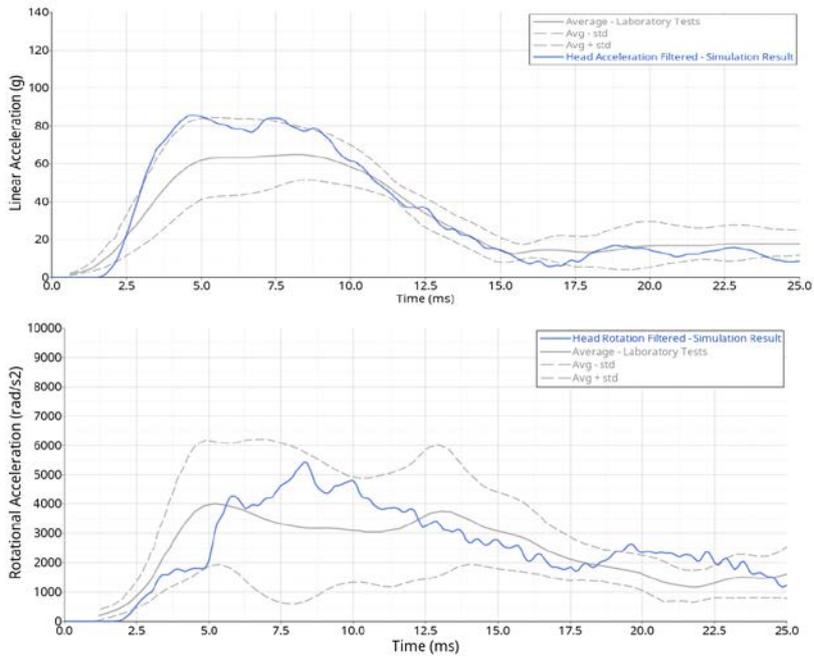

Figure 10: Case 77 - Struck player Hybrid III model response vs tests

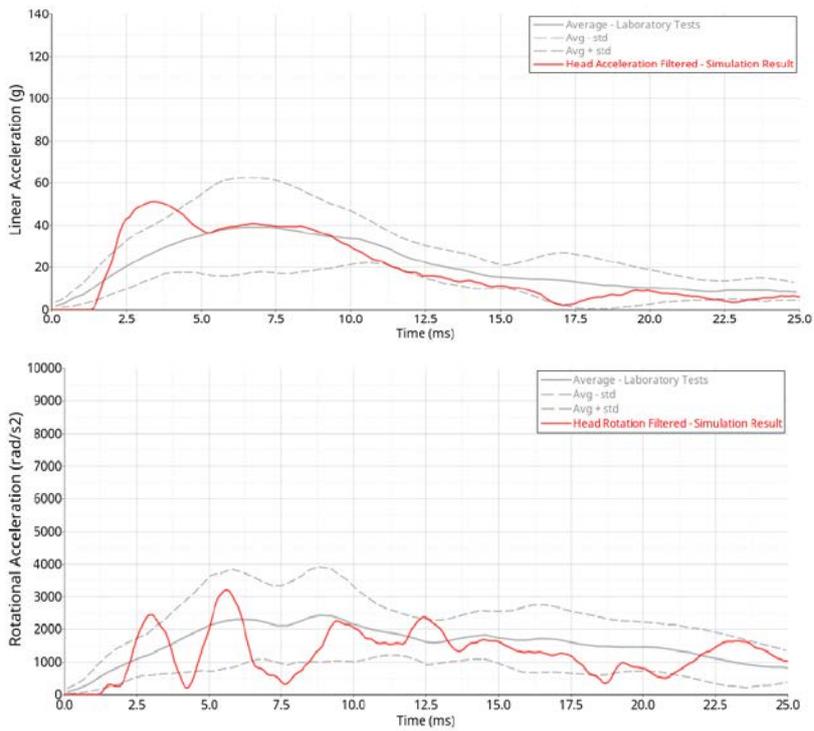

Figure 11: Case 77 - Striking player Hybrid III model response vs tests

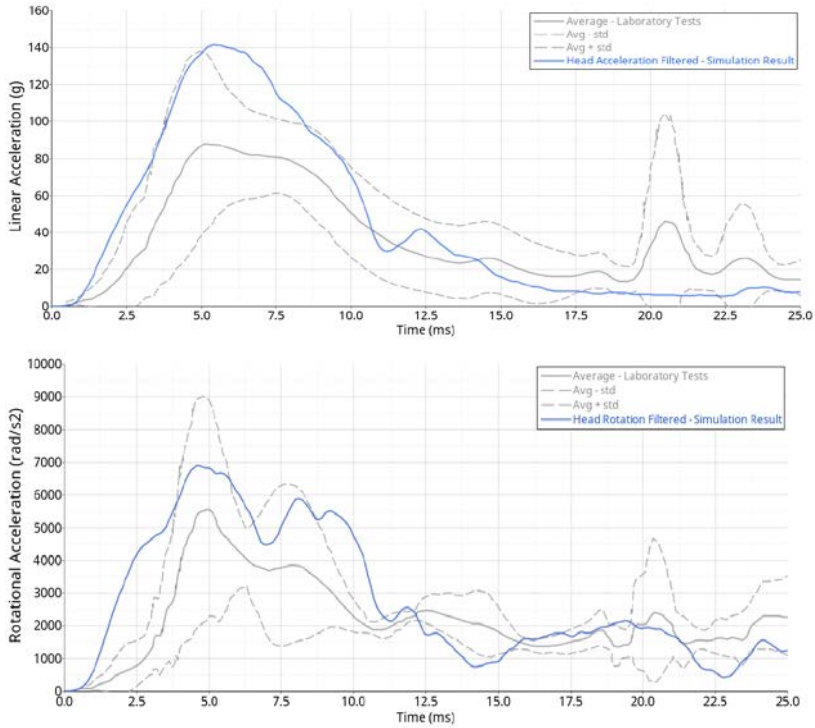

Figure 12: Case 164 - Struck player Hybrid III model response vs tests

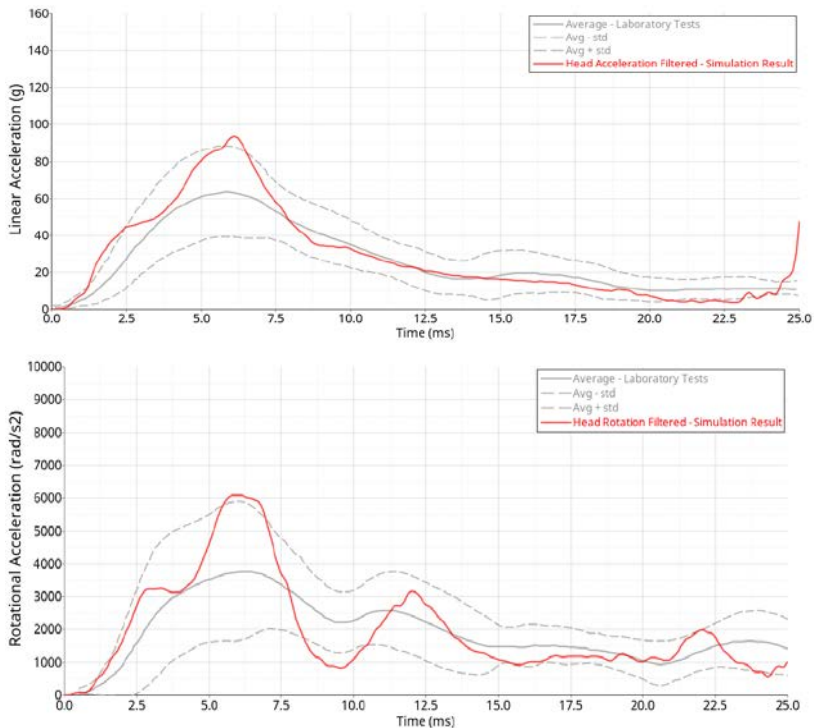

Figure 13: Case 164 - Striking player Hybrid III model response vs tests

All cases met the test corridors, highlighting that the impact boundary conditions are representative of the collisions which are being reconstructed.

## 3.2 PVP Results

Following the successful correlation, the Hybrid III headform was replaced by a THUMS 4.02 human head. The models were then recomputed and the PVP values for the centre of the brain for the struck and striking players were extracted as reported in Table 2.

|  |  |  | Impact Location | Impact Velocity (m/s) | PVP (mW) | Concussed (Y/N) |
|---|---|---|---|---|---|---|
| STRIKING | WM_CENTRE | CASE 38 | Lateral LHS | 9.5 | 0.928 | Y |
| STRUCK | WM_CENTRE | CASE 38 | Head Top | 0.0 | 0.951 | N |
| STRIKING | WM_CENTRE | CASE 77 | Head frontal | 9.9 | 9.405 | Y |
| STRUCK | WM_CENTRE | CASE 77 | Head Top | 0.0 | 0.243 | N |
| STRIKING | WM_CENTRE | CASE 164 | Lateral LHS | 10.8 | 2.923 | Y |
| STRUCK | WM_CENTRE | CASE 164 | Head Top | 0.0 | 1.184 | N |

Table 2: PVP results from the three collisions studied

## 4 Discussion

The three collisions' angular and linear accelerations measured in the FE impact simulation with the Hybrid III headform, meet the upper and lower corridor test responses. Consequently, the initial boundary conditions (impact velocities, impact angles, and locations) defined with the optimization process can be considered realistic and offer a solid foundation for studying brain distortion by replacing the Hybrid III headform with a THUMS 4.02 human model.

The test data used in this study are based from research in 2005 in the US. Consequently, the 'severity' of concussion is not known because the Glasgow Coma Scale (GCS) was not used there. As the players in the examined cases came back into the game, it can be reasonably assumed that any concussion recorded were mild, hence with a GCS between 15 and 13 (Table 3).

| GCS score | Ranking |
|---|---|
| 15-13 | Mild |
| 12-9 | Moderate |
| 8-3 | Severe |

Table 3: Glasgow Coma Scale [22]

As the three struck players were impacted on the top of the helmet and were not concussed (Figure 14), it is possible to propose a maximum allowable PVP value of 1.184mW to be measured in the white matter centre of the brain. This value can be used to design future safety protection gear or assess contact sports safety when subjected to vertical impacts.

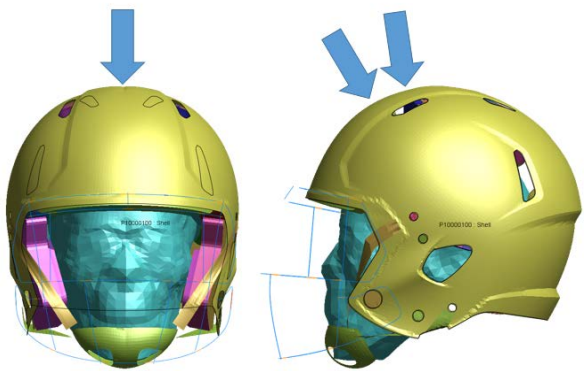

Figure 14: Proposed areas of helmet where concussion is avoided when PVP values < 1.184mW.

For the frontal and side impact directions, the information provided suggest mild concussion. Consequently, to avoid moderate and severe concussions, the PVP on side impact must be less than 0.928mW and for frontal impact 9.405mW. More information is necessary to extract the PVP values to avoid any concussions in these two directions.

The dataset used in this study, found in literature, was limited, hence future work would require international collaborations to include more cases, including a wide range of concussion levels, to define the PVP threshold for concussion in all head impact direction.

## 5  Conclusion

The research has evidenced that Peak Virtual Power (PVP) has the possibility to assess the concussion level, this concussion level varying depending on the impact direction. This was achieved by using previous research and applying the PVP method which extracted plausible maximum brain distortion segment locations, validated using imaging.

As an application of PVP, as plausible concussion computation method, the research reconstructed 3 NFL helmet-to-helmet impacts by means of finite element and using a validated Biocore helmets fitted with calibrated Hybrid III headforms. An optimisation routine extracted for each scenarios the helmeted head impact angles to correlate against linear and angular acceleration impact corridors defined as realistic impact conditions. Once the exact impact conditions are known, the Hybrid III headform was replaced by a validated THUMS 4.02 human head model in which the PVP was computed for each head, at the centre of the brain.

The research has concluded that in the vertical impact direction, no concussion would occur for a PVP value less than 1.184mW and that moderate and severe concussions could be avoided for side impacts and frontal impacts with PVP values lower than 0.928mW and 9.405mW respectively. With this innovative approach, it is possible to design safer head gear, as well as assess the safety of contact sports.

## 6  Limitations and Future Work

In this paper, it has been evidenced that PVP has the potential to investigate concussion. As future work it is proposed to investigate whether a relationship between PVP and the GCS can be established. A regression analysis could be used to estimate the values of the parameters. The format is expected to be of the shape of Equation 2.

$$PVP = PVP_T + K(15 - \text{GCS})^n$$

Equation 2: Proposed regression between PVP and GCS

Where:

- $PVP_T$ is the PVP threshold for a GCS value of 3, i.e. the lowest value for concussion
- K is a constant of proportionality
- 15 is the value of the GCS for no concussion
- n is an exponent

The dataset used in this study, found in literature, was limited, hence future work would require international collaborations to include more cases including a wide range of concussion levels to test the validity of Equation 2.